# Room-temperature insulating ferromagnetic (Ni,Co)$_{1+2x}$Ti$_{1-x}$O$_3$ thin films

Yukari Fujioka[1], Johannes Frantti*[1], Christopher Rouleau[2], Alexander Puretzky[2], Zheng Gai[2], Nickolay Lavrik[2], Andreas Herklotz[3,4], Ilia N. Ivanov[2] and Harry M. Meyer[2]

[1]Finnish Research and Engineering, Helsinki 00180, Finland

[2]Center for Nanophase Materials Sciences, Oak Ridge National Laboratory, Oak Ridge, Tennessee 37831, USA

[3]Materials Science and Technology Division, Oak Ridge National Laboratory, Oak Ridge, Tennessee 37831, USA

[4]Institute of Physics, Martin-Luther-University Halle-Wittenberg, Halle 06099, Germany

*Corresponding author: e-mail johannes.frantti@fre.fi

## ABSTRACT

Insulating uniaxial room-temperature ferromagnets are a prerequisite for commonplace spin wave-based devices, the obstacle in contemporary ferromagnets being the coupling of ferromagnetism with large conductivity. We show that the uniaxial $A_{1+2x}$Ti$^{4+}_{1-x}$O$_3$ (ATO), $A$=Ni$^{2+}$,Co$^{2+}$ and 0.6<$x$≤1, thin films are electrically insulating ferromagnets already at room-temperature. The octahedra network of the ATO and the corundum and ilmenite structures are the same yet different octahedra-filling proved to be a route to switch from the antiferromagnetic to ferromagnetic regime. Octahedra can continuously be filled up to $x$=1, or vacated (-0.24<$x$<0) in the ATO structure. TiO-layers, which separate the ferromagnetic (Ni,Co)O-layers and intermediate the antiferromagnetic coupling between the ferromagnetic layers in the NiTiO$_3$ and CoTiO$_3$ ilmenites, can continuously be replaced by (Ni,Co)O-layers to convert the ATO-films to ferromagnetic insulator with abundant direct cation interactions.

# INTRODUCTION

Magnetic, notably conducting ferro- and ferrimagnetic, materials are commonplace in modern microelectronics devices. Ferromagnetic insulators – still being sparse - have apparent advantages over the conducting magnetic materials. Oxides of magnetic cations are a natural starting point to look for ferromagnetic insulators, though they commonly order antiferromagnetically, as dictated by the oxygen mediated superexchange between the cations [1]. Ferromagnetism due to double exchange interaction was treated in ref. 2, which correlates the ferromagnetism and electric conductivity in $(La_{1-x}A_x)MnO_3$, $A$ is Ca, Sr, or Ba [3,4]. The double exchange interaction explains the correspondence between the largest magnetization and largest electrical conductivity. Cobalt modified $TiO_2$ anatase has been under keen interest due its ferromagnetic characteristics, though the observed ferromagnetism and electric conductivity seem to depend on the oxygen content and the origin of the ferromagnetism is not yet fully clear [5]. The subject of the present report is to show how ferromagnetic insulators can be formed by altering the strength and sign of the indirect and direct interactions, doable in the corundum and ilmenite-derived crystal structures. We also show how magnetization can be adjusted by adjusting composition, and briefly address potential thin film and interface applications.

Though insulating ferromagnets are rare, electrically insulating oxides with net magnetization due to uncompensated magnetization from the antiparallel alignment of magnetic sublattices are known. Also in these materials the ordering results in from antiferromagnetic interactions. Prime examples are ferrites, with either the spinel, garnet or hexaferrite structure, commonly applied in the RF-devices as films with thickness often exceeding 100 microns. Magnetic thin films find applications in miniaturized RF-devices utilizing spin waves as information carriers [6,7]. Miniaturization is not solely based on the smaller dimensions of individual devices but also on the possibility to diminish the number of devices via tunability. By forming layer structures from functionally different materials devices for versatile applications can be prepared. Examples are magnetic sensor, memory, and spintronic devices [8,9] and tunable phase shifters [10]. Insulating oxides are often transparent to visible and ultraviolet light. In the recent study transparency of the $Co_3O_4$ nanoparticles covered by L- and D-cys amino acids, to circularly polarized light in the ultraviolet range was reversibly modulated by magnetic field [11]. Bulk $Co_3O_4$ is not optically active since it has the cubic spinel structure. The study demonstrated the chirality transfer from amino acids to the crystalline core of the $Co_3O_4$ nanoparticles, creating chiromagnetic nanoparticles. Light beam modulation by external magnetic field can be applied in magnetooptics.

Spintronics is an area in which devices based on insulating ferromagnetic thin films are intensively studied. A spin filter tunnel junction was envisaged in ref. 12, and a tunnel junction spin filtering device, utilizing insulating ferromagnetic EuS layers decoupled by an $Al_2O_3$ layer was demonstrated in ref. 13. Spin filtering characteristics were measured at 1 K, the Curie temperature of EuS being 16 K. Garnet structured ferrites, mainly $Y_3Fe_5O_{12}$ (YIG) and its derivatives, are commonly utilized in room-temperature applications based on spin-wave propagation [8]. Since iron is a major metallic impurity in semiconductor industry (see, for instance ref. 14), there is a demand for non-ferrous materials. Due to the cubic symmetry the magnetocrystalline anisotropy of YIG is modest, requiring an external magnetic field for practical devices. An external magnetic field is impractical for miniaturized devices. Instead, ferromagnetic materials with uniaxial anisotropy can be applied in devices requiring uniform magnetization. In magnetic tunnel junctions CoFeB is frequently applied, though it requires buffer layers (commonly heavy metals Mo or Ta) to preserve the correct structure and composition. Due to the electric conductivity, spin-waves quickly attenuate in metallic ferromagnets.

Exchange bias - a phenomenon in which an antiferromagnet unidirectionally pins the adjacent ferromagnetic layer - plays a crucial role in magnetic devices. The material choice and layer growth technique affects the magnitude and the possible reduction of the exchange bias upon subsequent field cycling. The reduction is called the training effect, which appears both in in-plane and out-of-plane magnetized ferromagnetic films. The latter is often referred to as perpendicular exchange bias and is suggested to have a crucial role for applications [15]. A common device consists of a ferromagnetic layer whose magnetization direction is pinned by the antiferromagnet, and a ferromagnetic layer whose magnetization direction can be realigned. The device forms a basis for memory cells and spin filters. Good interfaces are critical for proper device functionality [16]. Also the oxidation level affects magnetic properties [17]. Attachment of structurally similar layers diminishes defect, such as misfit dislocations or strain, densities.

We report on an insulating uniaxial $A_{1+2x}Ti_{1-x}O_3$ (ATO) thin films [18], deposited by alternating target pulsed laser deposition [19]. As shown below, in the ATO films the largest magnetization coincides with the largest resistivity, in contrast to the ferromagnetism due to the double exchange interactions. The ATO structure can be visualized as the ilmenite structure in which the octahedra are vacated ($x<0$) or filled ($x>0$). Oxygen atoms are arranged similarly in the ATO, ilmenite and corundum structures. Cations instead are arranged differently, described in the Supplemental Material. Origin of the electrical and magnetic properties in $NiTiO_3$ and $CoTiO_3$ ilmenites [20-23] (structurally similar to ATO with $x=0$) and $Ti_2O_3$ (bears similarity to ATO with $x\approx-1$), $V_2O_3$ and $\alpha$-$Fe_2O_3$ corundum compounds are first summarized to understand the predominance of antiferromagnetic ordering and the influence of 3$d$-orbital occupancy. Instead of the corundum structure, $Co_2O_3$ [24] and $Ni_2O_3$ [25] adopt the bixbyite structure at ambient conditions and thus do not suit for a direct comparison in the present case. $V_2O_3$ and $Ti_2O_3$ demonstrate the impact of direct orbital overlap interactions and disordered distribution of cations at the octahedra sites to electrical resistivity, which plays an important role in the present material.

## MAGNETISM AND CONDUCTIVITY IN THE ILMENITE AND CORUNDUM COMPOUNDS

The magnetic structure of the transition metal oxides depends on the direct orbital overlap of cations, besides oxygen mediated interactions. According to ref. 20, magnetic interactions between 3$d$-cations are divided to oxygen intermediated and direct orbital overlap interactions. Due to the short distance and the orientation of $t_{2g}$-orbital lobes, direct orbital overlap interactions are relevant when octahedra share an edge or a face, summarized in Supplemental Material. The sign of the interactions depends on geometrical factors and the occupation of $d$-orbitals.

### *ILMENITE COMPOUNDS*
**Antiferromagnetic ordering in NiTiO$_3$ and CoTiO$_3$**

The valence band edge of the $A$TiO$_3$ ilmenite compounds, where the $A$-cation is Ni, Co or Fe is derived from oxygen 2$s$ and 2$p$ states and the conduction band edge from cation 4$s$ and 4$p$ states [21]. The cation 3$d$ states are localized within the energy gap, and thus the magnetism due to the localized 3$d$-orbital states is described in terms of direct and indirect interactions. The compounds are electrically insulating, though the localized states can result in polaron conductivity at elevated temperatures.

NiTiO$_3$ and CoTiO$_3$ ilmenites undergo a phase transition from paramagnetic to antiferromagnetic phase at 23 and 38 K, respectively. The low-temperature antiferromagnetic structure of NiTiO$_3$ and CoTiO$_3$ ilmenites is explained by oxygen mediated 90° interactions in the Ni$^{2+}$-O (Co$^{2+}$-O) hexagonal basal $ab$-plane, and the interactions between the Ni$^{2+}$-O (Co$^{2+}$-O) planes which separated by diamagnetic Ti$^{4+}$-O layers. The 90° $ab$-plane interaction is dominated by the coupling of the $t_{2g}$ and $e_g$ orbitals by oxygen. In the case of the more than half filled $t_{2g}$ orbitals and half-filled $e_g$ orbitals the interaction is ferromagnetic, which is the case of $Ni^{2+}$ ($t_{2g}^6 e_g^2$) and $Co^{2+}$ ($t_{2g}^5 e_g^2$). The interactions between the layers perpendicular to the $c$-axis are cation-oxygen-oxygen-cation (cooc) type, and are antiferromagnetic [21]. The interactions are rather weak, as shown by the low Néel temperatures. Magnetic anisotropy dictates the spin direction in bulk crystals [21], which is in the $ab$-basal plane in $NiTiO_3$ and $CoTiO_3$.

## CORUNDUM COMPOUNDS
### Para- and antiferromagnetic ordering in $Ti_2O_3$ and $V_2O_3$

The electronic energy band structure of $V_2O_3$ [26] and $Ti_2O_3$ [26,27] corundum structures bear similarities and both exhibit a metal-insulator transition (MIT) related to the anomaly in the $c/a$ ratio when temperature is varied through the transition. The covalent bonding between cations occurs both parallel and perpendicular to the hexagonal $c$-axis [28]. Below the transition temperature the symmetry of $V_2O_3$ is lowered to monoclinic $I2/a$ [29]. In literature electronic-energy-band-structure computations refer to the trigonal symmetry. To explain the MIT, a band gap should open at the transition temperature. In trigonal crystals the three-fold degenerate $t_{2g}$ states are split to nondegenerate $a_{1g}$ and two-fold degenerate $e_g$ states. The simplest explanation offered for $Ti_2O_3$ is that a gap is opened between a full $a_{1s}$ and empty $e_g^\pi$ states with decreasing temperature due to the anomalously decreasing $c/a$ ratio. Band structure computations [26-28,30-32] indicate that the Fermi level is located within a band formed from Ti/V 3$d$ $t_{2g}$-orbital derived from the overlapping $a_{1g}$ and $e_g^\pi$ states (see particularly refs. 26-28), whereas the antibonding $a_{1g}$ and $e_g^\pi$ states do not overlap [26]. O2$p$-orbital derived states lay deeper in the valence band, and the Ti/V 3$d$ $e_g^\sigma$-orbital derived states are few eV above the conduction band edge. Due to the overlapping of partially filled $a_{1g}$ and $e_g^\pi$ states both $Ti_2O_3$ and $V_2O_3$ should be metallic [27]. To find an explanation for the discrepancy between experimentally observed insulating state and the band structure prediction valid for an ideal structure, the cases in which the V-site is vacant and the ideally vacant octahedron site is occupied by V were considered in ref. 32. While the density-of-states per V atom at Fermi level for the $V_{11}O_{18}$ (one V-site vacant) is increased by 40% over the reference $V_{12}O_{18}$, the corresponding vacancy-interstitial $V_{12}O_{18}$ bands is diminished by 30% [32]. Thus, cation disorder has dramatic effect on the transport properties. The band gap can also be introduced by taking electron correlation into account [33,34].

In the corundum structure the $\approx 135°$ cation-oxygen-cation (coc) interactions, shown in Fig. 1(a), between the adjacent $ab$-planes competes with the cooc-interactions.

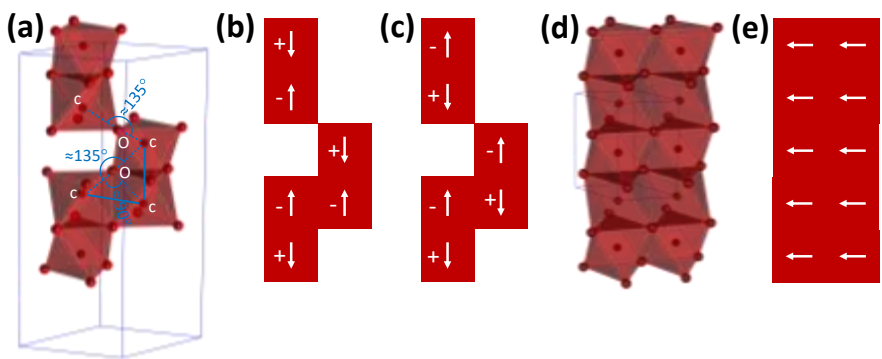

Fig. 1. Panel (a) shows the direct and oxygen intermediated interactions in the corundum structure. Panels (b) and (c) indicate two possible spin orderings in the corundum structure, respectively. The ordering +--+ shown in (b) is found in $\alpha$-$Fe_2O_3$ at low-temperatures and the ordering +-+- shown in (c) is found in $Cr_2O_3$. Panels (d) and (e) illustrate the ATO-structure for $x=1$, and magnetic ordering.

The latter interactions connect next neighbor layers. The coc-angle coupling the layers above and below is equivalent only if the octahedra are symmetric and the cation occupies the center. In the corundum structure the angles typically deviate by 10°. The angles depend on the lattice parameters and the fractional coordinates of the oxygen and cations within the unit cell. In compounds in which the $e_g$ orbitals are half filled the 135° coc-interaction is strong and antiferromagnetic. A prediction of the role of direct interactions in $Ti_2O_3$ and $V_2O_3$ is given in refs. 22 and 23. Based on the 3$d$ shell occupancies ($Ti3d^1$ and $V3d^2$), $Ti_2O_3$ and $V_2O_3$ were predicted to exhibit one and two transitions, respectively [22]. In $V_2O_3$, the high-temperature transition is due to the bond formation through the octahedra face along the $c$-axis, and the lower transition due to the bond formation through the octahedra edge in the $ab$-plane is discussed in refs. 22 and 23. $Ti_2O_3$ has only one broad transition in which Ti-cations form bonds parallel to the $c$-axis. Experiments are consistent with the bonding idea [23]. Straightforward indication of electron pairing is a reduced susceptibility and conductivity. With decreasing temperature the susceptibility of $V_2O_3$ sharply drops at around 180 K. A decrease in susceptibility is also seen in $Ti_2O_3$. The low-temperature phase of $V_2O_3$ is antiferromagnetic [35], though the details of the magnetic structure may be more complicated [36]. In contrast to $V_2O_3$, $Ti_2O_3$ exhibits no antiferromagnetic ordering [37] as the $3d^1$ electrons are paired at low-temperatures.

**α-Fe$_2$O$_3$ – a weak ferromagnet**

$Fe_2O_3$ stabilizes to the $\alpha$ and $\beta$ phases (the corundum and the bixbyite phases, respectively) at ambient conditions. The valence and conduction band edges of the α-$Fe_2O_3$ are essentially derived from the O 2$p$ states and Fe 3$d$ states, respectively, which predicts the compound to be of charge transfer type [38]. The low-conductivity of the α-$Fe_2O_3$ is due to the very heavy carrier masses in the conduction band minimum so that the conduction is polaronic. In α-$Fe_2O_3$ the 135° coc-interaction is stronger than the direct interaction and explains the magnetic order [21-23]. The layers perpendicular to the $c$-axis are ferromagnetically ordered. Basically, the direct cation-cation interactions favor antiferromagnetic ordering as do the 90° coc-interactions, analogously to $Mn^{2+}$ in $MnTiO_3$ [21]. However, the basal planes are ferromagnetic, since the strong 135° coc-interactions couple the ferromagnetic layer antiferromagnetically to the layer below and above. Thus, strong 135° coc-interactions explain the +--+ spin arrangement [39] along the $c$-axis, shown in Fig. 1(b). The magnetic ordering in α-$Fe_2O_3$ occurs without a change of the unit cell [40]. Due to the spin canting α-$Fe_2O_3$ exhibits weak ferromagnetism. In contrast, $Cr_2O_3$ exhibits +-+- spin arrangement [41] and the direct cation-cation interactions through shared octahedra face have dominant role, since the $e_g$ orbitals are empty.

*ATO* THIN FILMS – CHANGE IN MAGNETIC ORDERING VIA INCREASED DIRECT INTERACTIONS

In contrast to the ilmenite structure, the ATO structure of $(Ni,Co)_{1+2x}Ti_{1-x}O_3$ is no longer formed from alternating (Ni,Co)-O- and Ti-O-layers ($x$ = 0 being a possible exception), but the layers possess three types of cations. In the $x$ = 1 limit, all Ti-O-layers of the ilmenite structure are replaced by (Ni,Co)-O-layers and the number of cations in each layers is increased by 50%. The increase of the cation density is due to the complete filling of the octahedra, which affects the direct and oxygen mediated interactions between cations. In the ilmenite and corundum structures each octahedron forms a pair with the octahedron above (or below) along the $c$-axis direction via face sharing, and the pair has a vacant octahedron above and below. This favors direct interaction, and ultimately bonding, through the shared octahedron face. Filling of octahedra ($x$>0) corresponds to longer filled octahedra chains, which in turn affects the bonding, cation displacements and magnetism. For instance, the number of cations sharing octahedra faces both below and above increases with increasing $x$, which tends to displace the cation towards octahedron center. When $x$ = 0 or 1, the ATO structure possess a well-characterized space group symmetry, $R\bar{3}$ and $P6_3/mmc$, respectively. For other values of *x*, notably in the Ti-rich regime, the formation of local bonds expectably results in a non-cooperative phase transition spread over a broad temperature range. Figs. 1(d) and (e) illustrate the subject material of the present paper. By gradually filling the vacant octahedra sites by cations a ferromagnetic insulating phase is introduced. Ferromagnetism combined with electrical insulation is not obvious, as will be discussed.

# EXPERIMENTAL

Table 1 summarizes the samples addressed in this study. The compositions are given in terms of formula $A_{1+2x}Ti_{1-x}O_3$. Sample naming convention was introduced in ref. 18, except for the sample M(0.03), and for consistency the same convention is applied here except that from now on we give (x) after the sample name. The deposition procedure and results of the structural characterization are given in Table 1 in ref. 18. Briefly, $NiTiO_3$, $CoTiO_3$ and $NiCoO_4$ targets were repeatedly ablated, different substrate in-situ heating temperatures between 650 and 900K were tested, and the oxygen partial pressure was kept low, at 5mTorr in the case of sample A(-0.24) and at 10mTorr for the other samples reported in this study. Laser beam fluence was 3.0Jcm$^{-2}$, except the sample A(-0.24) which was grown by 4.8 Jcm$^{-2}$ fluence. Film thickness estimations are based on x-ray diffraction diffraction (XRD) and X-ray photoelectron spectroscopy (XPS) measurements. Homogeneous cation distribution in the films was confirmed by repeating a cycle in which a thin layer of film was sputtered and XPS spectra were subsequently collected until substrate was reached. This also yielded a thickness estimate of the films.

Table 1. The composition, substrate, thickness and fraction of the filled octahedra $z_v$. Except for the samples F(1) and I(0.61), the *c*-axis is perpendicular to the substrate plane. Thickness estimates in braces were obtained from XPS measurements. Ni, Co and Ti cation valences were +2, +2 and +4, respectively.

| Sample | Composition | Substrate | Thickness (nm) | $z_v$ |
|---|---|---|---|---|
| A(-0.24) | $(Ni_{0.57}Co_{0.43})_{0.51}Ti_{1.24}O_3$ | $Al_2O_3$ $(11\bar{2}0)$ | 53 (40) | 0.58 |
| C(0.24) | $(Ni_{0.39}Co_{0.61})_{1.49}Ti_{0.76}O_3$ | $Al_2O_3$ (0001) | 35.5 (34.6) | 0.75 |
| D(0.37) | $(Ni_{0.35}Co_{0.65})_{1.75}Ti_{0.63}O_3$ | $Al_2O_3$ (0001) | 39.0 (60) | 0.78 |
| E(1) | $(Ni_{0.39}Co_{0.61})_3O_3$ | $Al_2O_3$ (0001) | 44.3 | 1.00 |
| F(1) | $(Ni_{0.39}Co_{0.61})_3O_3$ | $Al_2O_3$ $(10\bar{1}0)$ | - | 1.00 |
| H(0.61) | $(Ni_{0.42}Co_{0.58})_{2.22}Ti_{0.39}O_3$ | $Al_2O_3$ (0001) | 53.7 | 0.87 |
| I(0.61) | $(Ni_{0.42}Co_{0.58})_{2.22}Ti_{0.39}O_3$ | $Al_2O_3$ $(10\bar{1}0)$ | - | 0.87 |
| J(0.03) | $(Ni_{0.46}Co_{0.54})_{1.05}Ti_{0.97}O_3$ | $Al_2O_3$ (0001) | 39.4 | 0.67 |
| M(0.03) | $(Ni_{0.46}Co_{0.54})_{1.05}Ti_{0.97}O_3$ | $Al_2O_3$ $(10\bar{1}0)$ | - | 0.67 |

Magnetization measurements were performed by a superconducting quantum interference device (SQUID) magnetometer (Quantum Design XL7). Field-cooled (FC) DC magnetization runs were carried out over the temperature range down to 5 K under 0.1 T magnetic field. Field-dependent magnetization data as a function of applied field were collected at 5 and 300 K. Data presented below were corrected for the substrate contribution by subtracting data measured from pure substrate under similar condition as the samples. Two measurement geometries were used, the applied field $H_a$ was either parallel (referred to as in-plane) or perpendicular (referred to as out-of-plane) to the substrate plane. Demagnetization field correction was applied for the out-of-plane data according to the equation $H_i=H_a-4\pi M$, where $H_i$ is the field inside of the film and $M$ is the magnetization in units of emucm$^{-3}$.

For electrical measurements interdigitated gold electrodes were patterned by photolithography and a mask. Electrodes were aligned differently in order to see possible orientation dependent conductivity.

# RESULTS

*FC-MAGNETIZATION AND HYSTERESIS*

Figure 2 shows the magnetization measured with constant magnetic field of 1000 Oe (FC-magnetization) as a function of temperature. Low-temperature anomalies are observed in all samples except to sample H(0.61). Sample J(0.03) has two-low temperature magnetic transitions. The humps seen in samples A(-0.24), C(0.24), D(0.37) and E(1) are weak and broad. None of the samples exhibits a curve typical to a pure antiferromagnetic order. Sample J(0.03), as shown below, exhibits co-existing antiferromagnetism and ferromagnetism at low-temperatures, and a clear connection between the hysteresis and FC-magnetization measurements. This is not inconsistent with the single phase nature of the films as discussed below. Weak antiferromagnetic feature is also seen in the sample E(1) at 5K.

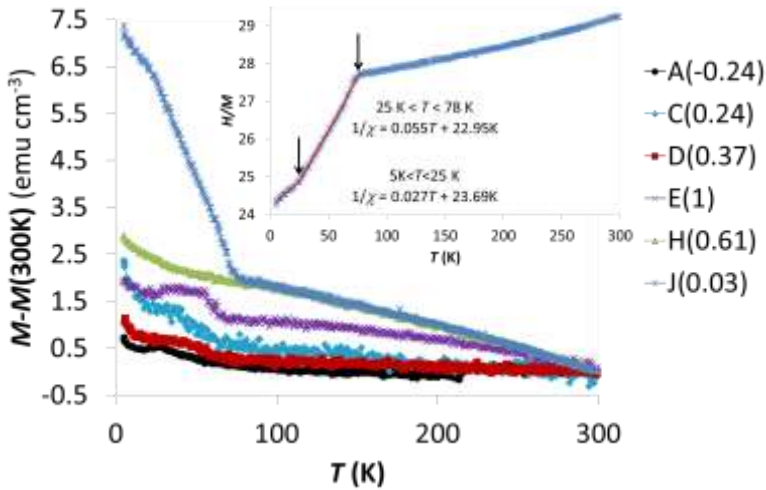

Fig. 2. FC magnetization as a function of temperature. Data were collected with a constant magnetic field 0.1 T with in-plane geometry. Room-temperature magnetization $M(300K)$ was subtracted to each curve, except for the curve shown in inset. The inset shows the $H/M$ curve of the sample J(0.03). Arrows indicate two transitions, one occurring at 78 K and the other one at around 25 K. The lower temperature transition was revealed by a slight change in the slope. Fits to two data ranges are given in the inset.

In sample E(1), the covalent bonding related to the $t_{2g}$ orbital overlap should first occur through the octahedra face along the *c*-axis, as it is the shortest direct distance between cations, see Table S1 in Supplemental Material. However, the anomaly observed in sample E(1) between 30 and 50 K does not correspond to a noticeable change in the hysteresis loop: $M$ increases slightly with decreasing temperature and thus, the anomaly is not related to a bond formation. Assuming that not all magnetic moments are ordered in high temperature region, the anomaly can be related to an ordering of the remaining moments. Another noticeable feature of the E(1) sample is that the magnetization is weaker than in the samples possessing less Ni and Co, such as the sample J(0.03). Also the difference between the in-plane and out-of-plane magnetization is small, though the magnetic moments lay in the *ab*-plane.

An indication of the role of the direct orbital overlap interactions is seen in Ti-rich sample A(-0.24) which becomes ferromagnetic only at low-temperatures. Within the direct interaction scheme this suggests that the hybrid state of the Co $t_{2g}$ and Ti $t_{2g}$ is half-filled by an electron originating from the half-filled Co $t_{2g}$ orbital. The appearance of ferromagnetism is consistent with the idea that the anomaly is not related to a covalent bond formation. Structural distortions, present already at room temperature, are observed also in the bulk $Ni_{1-x}Co_xTiO_3$ Solid Solution [42,43]. In thin films [18], similar distortions appear in the ilmenite-like sample J(0.03).

Fig. 3 shows the room-temperature and 5 K magnetization as a function of field, *M-H*, loops for in-plane geometry. Fig. 4 plots the out-of-plane *M-H* loops. Extracted values and the interpretations of the magnetic data are summarized in Table 2.

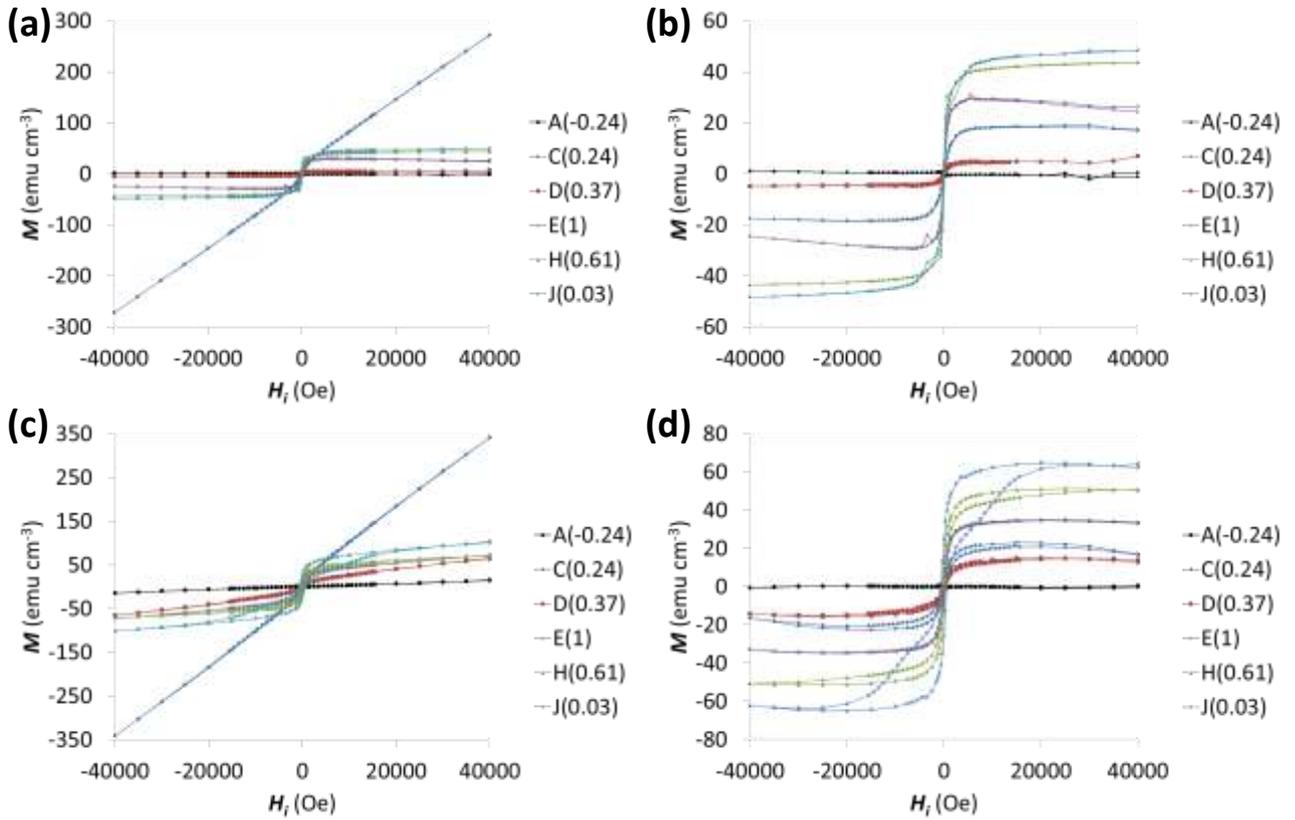

Fig. 3. In-plane *M-H* loops measured at 300 K [panels (a) and (b)] and 5 K [panels (c) and (d)]. Panels (b) and (d) give the linear part subtracted *M-H* loops.

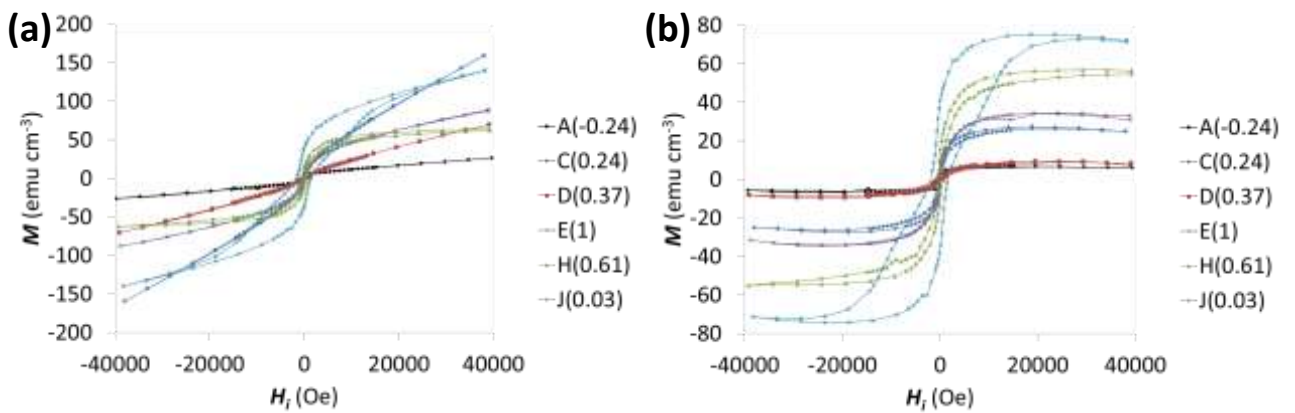

Fig. 4. Out-of-plane *M-H* loops measured at 5 K. Panel (a) shows the data including paramagnetic part, and panel (b) gives the linear part subtracted *M-H* loop.

Sample A(-0.24) is paramagnetic at room temperature. Except for the sample C(0.24), the linear part corresponding paramagnetic contribution is weak in the samples. The paramagnetic contribution of the sample C(0.24) is larger in in-plane geometry, compare Figs. 3(c) and 4(a). The curves from which the linear part is subtracted, Fig. 3(b) and (d), show that, besides sample A(-0.24), all samples possess a remnant magnetization at room-temperature. In the case of sample E(1), we note that $M$ slightly decreases with increasing $H$ at 300K, Fig. 3(b). This behavior is absent at 5 K. We assign it to an "overcorrected" substrate contribution. Similar behavior is seen in sample C(0.24) at 5K, Fig. 3(d).

Table 2. Magnetic properties as a function of composition and temperature. Abbreviations are: AFM – Antiferromagnetic, FM – Ferromagnetic, FiM – Ferrimagnetic, PM – Paramagnetic, $H_c$ – coercive field (in Oe), $M_r$ – remanence magnetization (emucm$^{-3}$), $M_s$ – saturation magnetization (emucm$^{-3}$), $M(T)$ – magnetization as a function of temperature. Labelings ∥ and ⊥ refer to the easy axis directions parallel and perpendicular to the *c*-axis direction, respectively, and coc(90°), cooc, cc(e) and cc(f) refer to 90° basal plane cation-oxygen-cation, cation-oxygen-oxygen-cation interactions, and interactions through the shared edge and face of octahedra, respectively.

| Sample | In-plane $M(T)$ | In-plane magnetic state at 300K, $H_c$, $M_r$, $M_s$ | In-plane magnetic state at 5K, $H_c$, $M_r$, $M_s$ | Out-of-plane magnetic state at 5K, $H_c$, $M_r$, $M_s$ | Easy axis direction | Dominating interactions |
|---|---|---|---|---|---|---|
| A(-0.24) | Weak and broad peak with a maximum at around 30K | PM | PM | FM, 170, 2, 6 | ∥ | cc(f) |
| C(0.24) | Weak and broad peak with a maximum at ≈40K | PM+FM, 131, 2, 19 | PM+FM, 250, 5, 23 | PM+FM, 175, 5, 25 | ∥ | cc(f), coc(90°), cc(e) |
| D(0.37) | Broad peak onset at 70K with a maximum at ≈40K | FM, 153, 1, 5 | FM, 210, 1, 14 | FM, 490, 2, 8 | ⊥ | cc(f), coc(90°), cc(e) |
| E(1) | Broad peak(s) at around 35 and 52K | FM, 99, 4, 28 | FM, 230, 8, 33 | FM, 291, 5, 32 | ⊥ | cc(f), cc(e), coc(90°) |
| H(0.61) | No anomalies | FM, 93, 5, 43 | FiM, 262, 12, 50 | FiM, 386, 12, 56 | ⊥ | cc(f), cc(e), coc(90°),cooc |
| J(0.03) | Slope change at 78K and 25K | FiM, 66, 5, 47 | FiM, 667, 25, 64 | FiM, 1867, 38, 72 | ⊥ | cc(f), cc(e), coc(90°) cooc |

*EASY AXIS AND COERCIVE FIELD*

All samples have the *c*-axis perpendicular to the substrate which provides information about the easy axis direction. As is seen by comparing Figs. 3(d) and 4(b), magnetization reaches saturation quicker in in-plane geometry. Clearest example is exhibited by the susceptibility curve of the sample J(0.03), which follows a curve characteristic to magnetic moments parallel to the substrate plane, shown in the inset of Fig. 2. Sample J(0.03) exhibits a double loop at 5 K, showing that the film is ferrimagnetic. FC curve, Fig. 2, reveals that this sample undergoes two phase transitions at 78 and 25 K. The composition and the structure of the sample J(0.03) is close to the prototype ilmenite, and its properties can be contrasted to the properties found in bulk materials. In the bulk $(Ni_{0.60}Co_{0.40})TiO_3$ there are two transitions, at around 25 and 69 K [42], the lower transition corresponding to a transition to ferrimagnetic phase with FC curves characteristic to an antiferromagnetic cusp. *M-H* measurements showed that the $(Ni_{0.60}Co_{0.40})TiO_3$ sample possesses a remnant magnetization at temperatures below 69 K [42]. The composition of the sample J(0.03) is similar to the bulk sample, and the FC curve is no longer of simple antiferromagnetic type. In contrast to the bulk sample, sample J(0.03) is ferromagnetic at room temperature. The magnetic moments are parallel to ferromagnetic layers and substrate plane, as can be concluded from the data shown in Fig. 2. The layers perpendicular to the *c*-axis are ferromagnetically coupled at room-temperature. A clear onset of antiferromagnetic coupling initiates at around 78 K. In this model the inverse susceptibility below 78 K is $1/\chi_\parallel$. The model is given in simplest terms and does not describe finer details, such as the change in the inverse susceptibility slope at around 25 K, seen also in the bulk sample. The difference with the phenomenological model given for uniaxial antiferromagnet in which the transition takes place between paramagnetic and antiferromagnetic phases (see, e.g. ref. 40, p. 124) is that the present transition occurs between phases in which the coupling between the ferromagnetic layers changes as a function of temperature. A clear double loop is seen at 5 K, Fig. 3(d) and 4(b), indicative of a significantly large fraction of antiferromagnetically coupled layers.

Coercive fields are smaller in in-plane case (see Table 2), the exception being sample C(0.24). This behavior can be explained by assuming that the easy axis is perpendicular to the *c*-axis in all samples except samples A(-0.24) and C(0.24). The easy axis of the samples A(-0.24) and C(0.24) is assigned to be the *c*-axis. Sample A(-0.24) is the most Ti-rich sample, which shows a weak but evident out-of-plane remnant magnetization at 5 K. No in-plane magnetization was found for sample A(-0.24) at 5 and 300 K, see Figs. 3(b) and (d). It is worth to note that though the sample A(-0.24) was grown on $(1\bar{1}20)$ substrate, the film exhibited preferred 00*l* orientation [18]. Shape anisotropy favors magnetization ***M*** in the plane of the film and often has a large effect on the direction of ***M***. The shape anisotropy due to the demagnetization energy affects the spin orientation and favors in-plane ***M***: the energy cost related to the orientation of ***M*** is proportional to $\cos 2\theta$, where $\theta$ is the angle between the film normal and ***M***. Despite shape anisotropy, the magnetization of the samples A(-0.24) and C(0.24) prefers to stay perpendicular to the substrate plane.

*SATURATION AND REMNANT MAGNETIZATION*
The saturation magnetization values extracted from the in-plane and out-of-plane geometries at 5 K are similar, except for sample D(0.37), see Table 2. The magnetization of the samples C(0.24), H(0.61) and E(1) saturated in both geometries. In sample D(0.37) magnetization parallel to the *c*-axis is not saturated, as is seen by comparing the $M_s$ values measured at 5 K for in-plane and out-of-plane geometries given in Table 2, consistently with the idea that the magnetic moments prefer to be in the *ab*-plane. The largest remnant and saturation magnetization values are found in sample J(0.03). Sample H(0.61) exhibits a weak double loop at 5 K, see Figs. 3(d) and 3(b). At room-temperature the sample exhibits remnant magnetization with a magnitude between the values measured from samples E(1) and J(0.03). Even though the density of Ni/Co cations is largest in the sample E(1), it had a saturation magnetization roughly half of the value found in sample J(0.03).

*ELECTRICAL RESISTANCE*
Table 3 lists specific resistance $\rho$ measured from thin films grown on $Al_2O_3$ $(10\bar{1}0)$ substrates.

Table 3. Specific resistance $\rho$, in M$\Omega$cm, measured at 1Hz frequency.

| Sample | $\rho$ |
|---|---|
| F(1) | 2.2 |
| I(0.61) | 135 |
| M(0.03) | 170 |

In terms of the hexagonal axes, only 220 reflection in the sample F(1), and 110 and 220 reflections in the sample I(0.61), were observed and thus the conductivity values correspond to the conductivity in the 110-plane (11$\bar{2}$0-plane in hexagonal indices). Despite the substrate orientation, sample M(0.03) had *c*-axis perpendicular to the substrate plane, and thus the orbital overlap interactions through octahedra edges are relevant for understanding the conductivity.

Measurements on interdigitated electrodes, oriented randomly on the sample surface, indicated no dependence on the crystal orientation on the plane. Among the three samples, $\rho$ increases with decreasing *x*, i.e., with decreasing octahedra filling fraction. Though all films are insulating, the nearly two orders of magnitude smaller $\rho$, when compared with the similarly oriented film I(0.61), found in film F(1) suggests that there is a dependency between the extent of direct cation overlap and increased conductivity. In sample F(1), every octahedra is filled by Ni or Co. Cation-cation interactions due to the overlapping $t_{2g}$ orbitals can result in formation of a fully occupied band.

*FERROMAGNETIC AND ANTIFERROMAGNETIC COUPLING*

In the ATO films the magnetic moments are ferromagnetically arranged in the *ab*-planes, as in NiTiO$_3$ and CoTiO$_3$ [21]. The 90° coc-interaction between Ni$^{2+}$ and Co$^{2+}$ cations remains ferromagnetic when the filling fraction is increased. In NiTiO$_3$ and CoTiO$_3$ the ferromagnetic *ab*-plane coupling cooperates with the antiferromagnetic cooc-interaction. The ferromagnetic order, stable at room-temperature in thin films, is related to the compressive *ab*-plane stress which enhances the 90° coc-interactions as the distances become shorter. Due to the difference between Al$_2$O$_3$ substrate lattice parameters and the film lattice parameters the *ab*-plane is under compressive stress, whereas the *c*-axis is stretched [18], thus the cooc-interaction becomes weaker. This, however, does not explain the ferromagnetic coupling between the *ab*-planes. Instead, we assign the coupling by cation-cation interactions.

In strained piezomagnetic films, magnetization has a term linearly proportional to the strain. Whether the film exhibits piezomagnetism depends on the magnetic point group symmetry. As an example, the magnetic point groups of the structures shown in Fig. 1(b) and (c) do not exhibit piezomagnetic effect, whereas the ++-- and ++++ arrangements correspond to the piezomagnetic symmetries [40]. The symmetry depends on the direction of the magnetic moments. Ferromagnetic phases (phases with different magnetic moment directions counted distinct) of the *x*=1 films are piezomagnetic when the magnetic moments are directed along the *x*- or *y*-axis [44]. The magnetic point group for both cases is $2/m\,2/m'\,2/m'$. In this case, the magnetization depends linearly on stress. In contrast, if the magnetic moments are parallel to the *z*-axis, the magnetic point group is $6/m\,2'/m'\,2'/m'$, which does not allow piezomagnetic effect. Table 2 shows that magnetization does not monotonically increase with decreasing *x*, which could be due to different strain in films.

To understand the ferromagnetic coupling along the *c*-axis, we consider the role of direct cation-cation interactions. Degeneracy of the $t_{2g}$ states is a valid assumption for Ni and Co, which have 8 and 7 *d*-electrons, respectively. The Co-Co direct orbital interactions through shared octahedra face and edge are ferromagnetic [23], consistently with decreasing antiferromagnetic coupling with decreasing Ti-content. Also the correlation superexchange mechanism for 90° coc-interaction is ferromagnetic for interacting Ni/Co cations [20]. As illustrated in Supplemental Material, the layers adjacent along the *c*- axis are coupled via 71° coc-interactions. In the case the $t_{2g}$ orbitals are more than half-filled, two $p$ electrons from different $p$ orbitals ($p\sigma$ for each cation) of like spin are simultaneously excited from the oxygen ion to the $e_g$ orbitals of the coupled cations [20]. In contrast to the corundum structure, in samples E(1) and F(1) each cation has equivalent direct interaction with the cations above and below. No susceptibility decrease, indicative of a covalent bond formation, is seen in Fig. 2. As samples E(1) and J(0.03) show (Figs. 3 and 4), the competing 135° coc-interaction is still present and is sufficiently strong to couple significant fraction of the *ab*-planes antiferromagnetically at low-temperature. Further increase in the Ti-content decreases the magnetic cation content and magnetization. Though all octahedra sites can be filled by Co and Ni, pure cobalt oxide thin films, grown under similar conditions as (Ni,Co)-rich samples, stabilized to spinel structure [18].

A further contrast to the antiferromagnetic corundum and ilmenites compounds is that the ferromagnetic and antiferromagnetic interactions are not cooperative in the thin films. This is seen from the behavior of the samples H(0.61) and J(0.03): both are ferromagnetic at room-temperature, whereas the cooperative antiferromagnetic transition takes place at low-temperature in sample J(0.03). As Figs. 3 and 4 show, sample H(0.61) is ferrimagnetic at low-temperatures yet exhibits no anomaly in magnetization as a function of temperature, see Fig. 2. Thus, the antiferromagnetic ordering occurs gradually. The antiferromagnetic coupling does not correspond to a change in the ferromagnetic component. The ferromagnetic coupling occurs at higher temperature than the antiferromagnetic coupling also in the bulk ilmenite (Ni$_{0.60}$Co$_{0.40}$)TiO$_3$ [42], and the ferromagnetic magnetization is preserved at lowest temperatures. The magnetic phase of a single phase sample can be simultaneously antiferromagnetic and ferromagnetic, if the symmetry groups of the ferromagnetic and antiferromagnetic vectors are the same [40]. This is demonstrated in $\alpha$-Fe$_2$O$_3$, in which the magnetic moments are canted and result in the weak ferromagnetic phase via Dzyaloshinskii-Moriya interaction. In (Ni+Co)-rich samples, the lack of antiferromagnetic ordering implies that the net magnetization is not due to the Dzyaloshinskii-Moriya interaction, and the 135° coc-interactions no longer dominate.

## CONCLUDING REMARKS

Magnetic properties of (Ni,Co)$_{1+2x}$Ti$_{1-x}$O$_3$, -0.24<$x$<1, thin films with ATO structure were addressed. Increased octahedra-filling increases the direct orbital interactions between the cations, which strongly alters the electrical and magnetic properties. ATO films possess uniaxial magnetocrystalline anisotropy and the octahedra filling and magnetization can be controlled during film growth by adjusting $x$. Films with magnetization parallel and perpendicular to the substrate plane were grown. The magnetic ion density of the $x = 1$ films is 3-fold when compared to the NiTiO$_3$ or CoTiO$_3$. In contrast to the prototype ilmenite and corundum structures, prevailingly exhibiting low-temperature antiferromagnetic ordering or at best canted ferrimagnetic ordering, the ATO films with $x \approx 0$ were ferrimagnetic, and films with $x \gtrsim 0$ were ferromagnetic already at room-temperature. The sample with x≈0 exhibited a ferrimagnetic ordering, which was profound below 78 K with a clear antiferromagnetic ordering. The sample had a magnetization in the entire measurement temperature range between 5 and 300 K. Conductivity measurements indicated that the samples with $x\gtrsim$0 are insulating ferromagnetic materials.

## ACKNOWLEDGMENTS
All experimental work was conducted at the Center for Nanophase Materials Sciences, which is a DOE Office of Science User Facility.

# SUPPLEMENTARY MATERIALS

## $A^{2+}_{1+2x}Ti^{4+}_{1-x}O_3$ (ATO) STRUCTURE

The $A$TO structure possesses the same oxygen octahedra network as the ilmenite structure. The essential difference between the ilmenite and ATO structures is the filling fraction of the octahedra and the cation arrangement followed by considerable changes in the crystal symmetry and magnetism. The prototype $AB$O$_3$ ilmenite structure is layered with $A$O$_6$ and $B$O$_6$ octahedra alternating along the hexagonal $c$ axis. Each layer consists of two, slightly separated, two-dimensional triangular cation nets. Within the layers the $A$O$_6$ and $B$O$_6$ octahedra share edges, and out of the layers (along the hexagonal $c$ axis) the octahedra are connected via shared octahedral faces and corners. Each octahedron shares edges with three filled and three empty octahedra and a face with one filled octahedron and one empty octahedron. This is the origin of the cation displacements towards the vacant octahedron. Two thirds of the octahedra are occupied by cations and the remaining one third of the octahedral-site vacancies are ordered so as to minimize the cation–sublattice electrostatic repulsive forces.

Figure S1(a) illustrates the $A$TO structure, obtained from the ilmenite structure by either filling (0<$x$1) or vacating (-½<$x$<0) the octahedra. The Ti/(Ni+Co) ratio is adjustable in the range -0.25 < $x$ < 1, and can be continuously changed during the growth cycle while preserving the crystal structure the same. The symmetry changes from rhombohedral to hexagonal (space group $P6_3/mmc$) at around $x$=0.61. The fraction of the filled octahedra, $z_v$, is a function of $x$, and the ilmenite and corundum structures are special cases of the $A$TO structure at which $x$ = 0. The dark-grey (TiO$_6$) and light-grey (Ni/CoO$_6$) octahedra indicate octahedra filled in the ilmenite structure. In a general case there is no particular ordering of the cations. The octahedra vacant in the ilmenite and the corundum structures are indicated in Fig. S1(a). The corundum structure is obtained by placing (often in an average sense) the same type of cations into the dark- and light-grey octahedra. When -½<$x$<0 each excess Ti$^{4+}$ (1) replaces a Ni/Co$^{2+}$ cation and (2) creates a vacant octahedron.

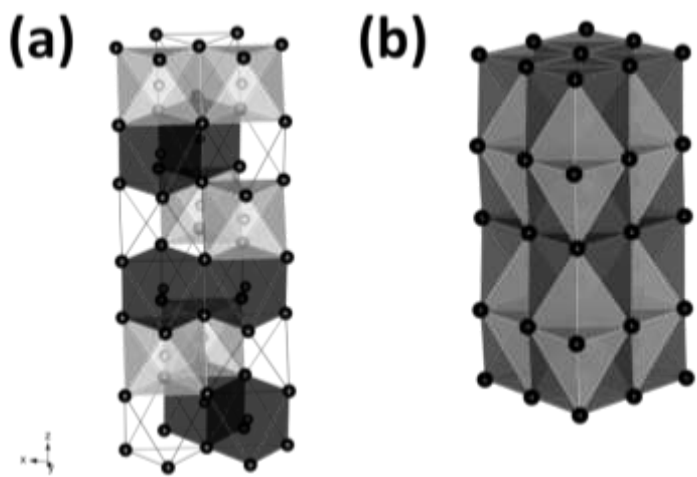

Fig. S1. (a) The $A$TO structure, given in terms of the hexagonal axes. Both the ilmenite and corundum structures correspond to the case $x$=0, $z_v$=2/3, the Ti-rich $A$TO structure to the case $x$<0, $z_v$<2/3 and the (Ni,Co)-rich $A$TO structure to the case $x$>0, 2/3≤$z_v$≤1. (b) The ATO structure when all octahedra (dark grey) are filled. The light-grey polyhedra are formed from two face-to-face tetrahedra and are predominantly empty.

In practice the Ti valence changes at high Ti concentrations - titanium is in a valence state +3 in the Ti$_2$O$_3$ corundum structure. If 0<$x$<1, two excess Ni/Co$^{2+}$ cations are required to replace one Ti$^{4+}$: the first excess Ni/Co$^{2+}$ replaces a Ti$^{4+}$ cation and the second one occupies a vacant octahedron. When $x<0$ less than 2/3 of the octahedra are filled. The octahedra filling mechanisms are consistent with the charge neutrality requirement and experimentally verified compositions and cation valences [17]. The $A$TO structure allows a large variety of cation arrangements.

As shown in Fig. S1(b), the ATO structure also contains chains of tetrahedra, connected via alternating vertex and face-sharing. No indication of cations in the tetrahedral sites was found [18], though it is possible that some of the sites are occupied. When occupied the distance between the nearest neighbor octahedron and tetrahedron cations through a shared face is particularly short, 1.7Å. If the tetrahedral sites would be occupied, there would be 180° cation-oxygen-cation interactions along the $c$-axis.

## OCTAHEDRA SHARING

Magnetic ordering is controlled by the type of cations and the octahedra filling configuration. Magnetization and electric dipole moment depend on octahedra filling fraction $z_v$. The statistics of the cation-cation interaction via shared octahedra face and edge changes as a function of $z_v$. In the prototype ilmenite and corundum structures each octahedron shares one face with one octahedron and for each filled octahedron the number of octahedra sharing just one face is decreased by two. When $x = 1$, every octahedron shares two faces, which are perpendicular to the $c$-axis.

In the $A$TO structure cations are displaced towards the octahedra centers with increasing $z_v$. When $x = 1$, each oxygen is shared by six octahedra and each octahedron is connected to 20 octahedra. Connection is via one (shared corner), two (shared edge) or three oxygen (shared face), as illustrated in Fig. S2. Within the basal plane each octahedron has six edge sharing neighboring octahedra (e.g., octahedron 1 shares an edge with octahedra 2 and 3), and above (similarly below) it has six corner sharing octahedra (e.g., octahedron 1 shares a corner with octahedra 5 and 6) and two face sharing neighbors (e.g., octahedron 1 shares a face with octahedron 4). Table S1 lists the corresponding cation-oxygen-cation angles and bond lengths. The bond lengths and angles depend on structural parameters, which change as a function of $z_v$, and strain due to the lattice mismatch between the substrate and thin film.

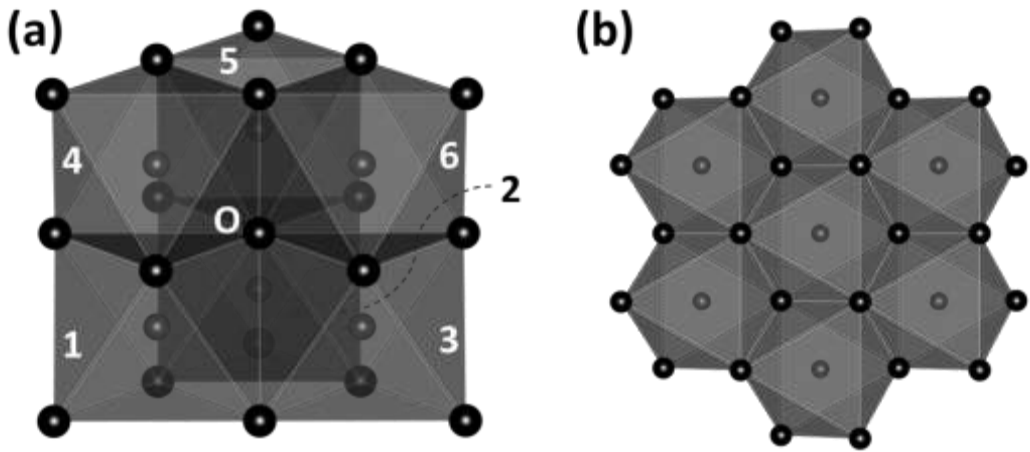

Fig. S2. (a) The six octahedra referred to in Table S1. Numbers refer to $i$ and $j$ used in Table S1. Left-hand picture illustrates the face, edge and corner sharing octahedra. (b) Right-hand picture shows that in the $ab$-basal plane each octahedron shares each edge with a neighboring octahedron.

Table S1. Cation-oxygen-cation angles and cation-cation distances in the $x = 1$ $A$TO structure. C, E and F refer to Corner, Edge and Face sharing of octahedra pairs $i$-$j$ specified in Fig. S2, respectively. $A_i$ and $A_j$

refer to cations in octahedra *i* and *j*. Angles and distances are computed for space group $P6_3/mmc$ with $a$ = 2.9780 Å and $c$ = 4.8703 Å. cc and coc refer to direct cation-cation overlap and cation-oxygen-cation interactions, respectively. Interactions expected to be dominant are given for each pair. The expectation is based on the known ordering found in the 3*d*-transition metal ilmenite and corundum oxides.

| *i-j* | Sharing | $A_i$-O-$A_j$ angle (°) | $A_i$-$A_i$ distance (Å) | Interactions |
|---|---|---|---|---|
| 3-6 | F | 70.6 | 2.44 | cc |
| 1-4 | F | 70.6 | 2.44 | cc |
| 2-5 | F | 70.6 | 2.44 | cc |
| 1-3 | E | 89.9 | 2.98 | cc, coc |
| 4-6 | E | 89.9 | 2.98 | cc, coc |
| 1-2 | E | 89.9 | 2.98 | cc, coc |
| 2-3 | E | 89.9 | 2.98 | cc, coc |
| 4-5 | E | 89.9 | 2.98 | cc, coc |
| 5-6 | E | 89.9 | 2.98 | cc, coc |
| 1-5 | C | 131.8 | 3.85 | coc |
| 1-6 | C | 131.8 | 3.85 | coc |
| 2-4 | C | 131.8 | 3.85 | coc |
| 2-6 | C | 131.8 | 3.85 | coc |
| 3-4 | C | 131.8 | 3.85 | coc |
| 3-5 | C | 131.8 | 3.85 | coc |